# Geo-Spatial Cluster based Hybrid Spatio-Temporal Copula Interpolation

Debjoy Thakur[1], Ishapathik Das[1]

[1] Department of Mathematics and Statistics, Indian Institute of Technology, Tirupati, India.

corresponding-author- debjoythakur95@gmail.com

**Abstract.** In the absence of Gaussianity assumptions without disturbing spatial continuity interpolating along the whole spatial surface for different time lags is challenging. The past researchers pay enough attention to Spatio-temporal interpolation ignoring the dynamic behaviour of a spatial mean function, threshold distance, and direction of maintaining spatial continuity. Therefore, we employ hierarchical spatial clustering (HSC) to preserve local spatial stationarity. This research work introduces a hybrid extreme valued copula-based Spatio-temporal interpolation algorithm. Spatial dependence is captured by a blended extreme valued probability distribution (BEVD). Temporal dependency is modeled by the Bi-directional long short time memory (BLSTM) at different temporal granularities, 1 month, 2 months, and 3 months. Spatio-temporal dependence is modeled by the Gumbel-Hougaard copula (GH). We apply the proposed Spatio-temporal interpolation approach to the air pollution data (Outdoor Particulate Matter (PM) concentration) of Delhi, collected from the website of the Central Pollution Control Board, India as a crucial circumstantial study. This article describes a probabilistic-recurrent neural networking algorithm for Spatio-temporal interpolation. This Spatio-temporal hybrid copula interpolation algorithm outperforms and is efficient enough to detect spatial trends and temporal influence. From the entire research, we detect that PM concentration in a year reaches a maximum, generally in November and December. The northern and central part of Delhi is the most sensitive regarding air pollution.

## 1 Introduction

In Geo-statistics interpolation along the entire spatial surface is interesting enough. Many scientists prefer neural networking-based Spatio-temporal interpolation like Geo-Long Short Time Memory (Geo-LSTM) (Ma J. a., 2019), Random Forest Regression Kriging (RFK), IDW-BLSTM (Ma J. a., 2019), and others. Nowadays, usage of copula namely, Gaussian copula, Factor copula (Krupskii, 2018), Archimedean copula (Sohrabian, 2021), Vine copula (Graler, 2014), (Shan, 2021), and so on in spatial interpolation is interesting enough. The previous researchers are reluctant about the Gaussianity assumption. In this research, maintaining spatial continuity, a new hybrid copula for Spatio-temporal interpolation we capture dynamicity satisfactorily.

## 2 Methods

We replace missing data with the predicted values of BLSTM, a generalization of LSTM architecture consisting of three different gates. (i) Forget gate ($F_t$) determines whether information of the last time point would be trimmed or carried out to the next step. (ii) Input Gate ($I_t$) decides whether the new information from the input will be added and the state of the memory cell at $t^{th}$ time point ($C_t$). (iii) Output Gate ($O_t$) identifies which information will be executed and formulated.

Let $\mathbf{S} = \{s_1, s_2, \dots, s_n\}$ and $\mathbf{T} = \{t_1, t_2, \dots, t_k\}$ are the spatial and temporal sample space. A random variable (RV) $Z: \mathbf{S} \times \mathbf{T} \mapsto \mathbf{R}^+$ is used to define an observation matrix

$$\mathbf{Z} = \begin{bmatrix} Z(s_1, t_1) & \cdots & Z(s_1, t_k) \\ \vdots & \ddots & \vdots \\ Z(s_n, t_1) & \cdots & Z(s_n, t_k) \end{bmatrix} \in \mathbf{R}^{n \times k}$$

Using $\mathbf{Z}$ a spatial influence ratio (SIR), $\text{SIR} = \frac{Z(s_{i_1}, t_j)}{Z(s_{i_2}, t_j)} = r_h \ni i_1 < i_2; \forall j = 1, 2, \dots, k$. Assuming $\mathbf{S}$ is isotropic and intrinsic stationary, spatial lag-dependence (SLD) is a validation range of SIR. $\text{SLD}: \Omega_h \mapsto \mathbf{R}^+$ and $\Omega_h$ is a sample space of $r_h$ such that $\text{SLD}(x) = \text{Max}\{\|h\| \subset \mathbf{R}^+ \ni r_h = x\}$. An empirical cumulative distribution function (CDF) of SLD is $\hat{F}(h) = P[\{r_h \in \Omega_h \ni \text{SLD}(r_h) \leq h\}]$. For flexibility we fit the blended extreme valued probability distribution (BEVD) (Vandeskog, 2022) i.e., $F(x) = (F_1(x))^{T(x; a, b)} * (F_2(x))^{1-T(x; a, b)}$ based upon maximum likelihood estimation (MLE) technique. Here $F_1(x), F_2(x)$ are the suitable EVD and the distortion function, $T(.)$ is the CDF of Kumaraswamy Distribution, $T(x; \theta = \{a, b \in \mathbf{R}^+\}) = F_{KS}(\frac{x-a}{b-a}; \theta) = 1 - (1 - x^a)^b$. It gives 1 unless it reaches $a$ and 0 after that. Similarly, temporal influence ratio (TIR) where, $\text{TIR} = r_\tau = \frac{Z(s_j, t_{i_1})}{Z(s_j, t_{i_2})} \ni i_1 < i_2$ and $\|t_{i_1} - t_{i_2}\| = \tau$. Temporal lag dependence function (TLD) where, $\text{TLD}: \Omega_\tau \mapsto \mathbf{R}^+$ and $\Omega_\tau$ is a sample space of $r_\tau$ such that, $\text{TLD}(y) = \text{Max}\{\|\tau\| \subset \mathbf{R}^+ \ni r_\tau = y\}$. The CDF of TLD is $\hat{G}(\tau) = P[\{r_\tau \in \Omega_\tau \ni \text{TLD}(r_\tau) \leq \tau\}]$. Again, we fit a suitable parametric EVD. Making use of spatial CDF, $F(h; \Theta_h)$, and temporal CDF, $G(\tau; \Theta_\tau)$ employing copula we create a bi-variate CDF (BCDF), $H(h, \tau) = C(F(h; \Theta_h), G(\tau; \Theta_\tau))$. From BCDF the bi-variate probability density function (BPDF) of STLD is, $f_{\text{SLD,TLD}}(u, v)$ where $\text{SLD}(x) = u$; $\text{TLD}(y) = v$. With the help of $f_{\text{SLD,TLD}}(u, v)$ we define a Spatio-Temporal interpolator in the following:

$$(h^*, \tau^*) = \arg\max_{(h, \tau)} \{f_{\text{SLD,TLD}}(u, v): \|h\| \leq u = \text{SLD}(r_h); \|\tau\| \leq v = \text{TLD}(r_\tau)\}$$

The above equation provides the most likely Spatio-Temporal lag vector $(h^*, \tau^*)$ corresponding to a specific Spatio-Temporal influence ratio ($\text{STIR} = (r_h^*, r_\tau^*)$). We introduce a mapping $a(.)$ from the vector space $\mathbf{M}$ (dimension $n_1$ over $\mathbf{R}^+$) of most likely Spatio-Temporal lag (MSTL) to the vector space of STIR, $\mathbf{P}$ of same dimension over same field. Here a: $\mathbf{M} \mapsto \mathbf{P} \ni$ a $(h_i, \tau_j) = (r_{h_i}, r_{\tau_j})$. The interpolated values at $(s_0, t_0)$ is $Z^{int}(s_0, t_0) = Z(s_i, t_j) * \sqrt{[\langle a(||s_i - s_0||, |t_j - t_0|), e_1 \rangle]^2 + [\langle a(||s_i - s_0||, |t_j - t_0|), e_2 \rangle]^2}$. Thus, we perform the interpolation and to measure accuracy we focus root mean square error (RMSE) and mean absolute error (MAE) of this model.

## 3   Results

We illustrate this method by taking the air pollution data of Delhi in 2019, collected from the CPCB website. We divide the study area into 4 HSCs with a radius of 18026m (Thakur, 2022). The most suitable distribution is, Weibull with shape parameters 4.8763, 0.5647 and scale parameters 1.829, 0.084 with loglikelihood value -29.816723. The RMSE and MAE are respectively 1.487335, and 0.4871.

## 4   Discussion

We validate the satisfactory performance of BLSTM neural network in the Fig. 1. In this figure we compare the performance of this neural network splitting the data set in two random parts tarin, and test respectively. We interpolate along the entire study area using spatial cluster-based hybrid Spatio-Temporal copula interpolation. Division of distinct spatial regions is as crucial as the variety of rolling window sizes so that spatial and temporal influences are not confounded. In Fig. 2 we depict the Spatio-Temporal dynamicity of monthly $PM_{2.5}$ emission in Delhi in the year, 2019 in every disjoint spatial region. We observe that the monthly $PM_{2.5}$ emission during April is varying 40-75μg/m$^3$, in May 50-100μg/m$^3$ whereas, in August that decreases and is limited to the interval of 20-40μg/m$^3$. Therefore, we can conclude that the emission is generally under control during summer and monsoon. But from the month of October, it starts to increase and ranges within 120-170μg/m$^3$. During winter this $PM_{2.5}$ ejection becomes very high especially, in the month of November differs from 180-220μg/m$^3$ again it decreases in December. We detect that air pollution mostly happens in the northern and central parts of Delhi.

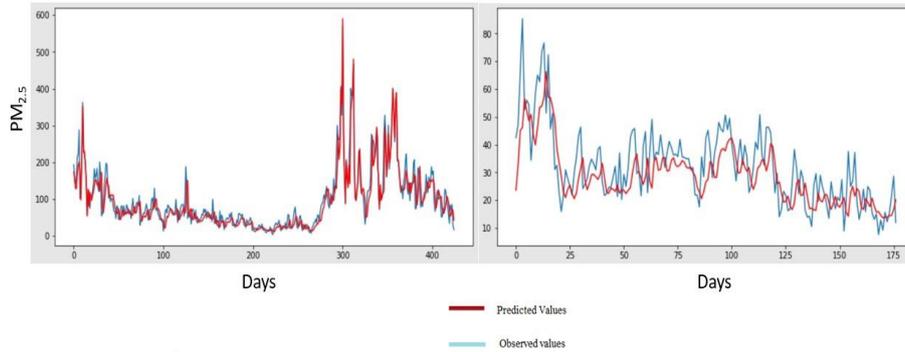

**Fig. 1.** Hybrid Spatio-temporal copula interpolation for different rolling window size.

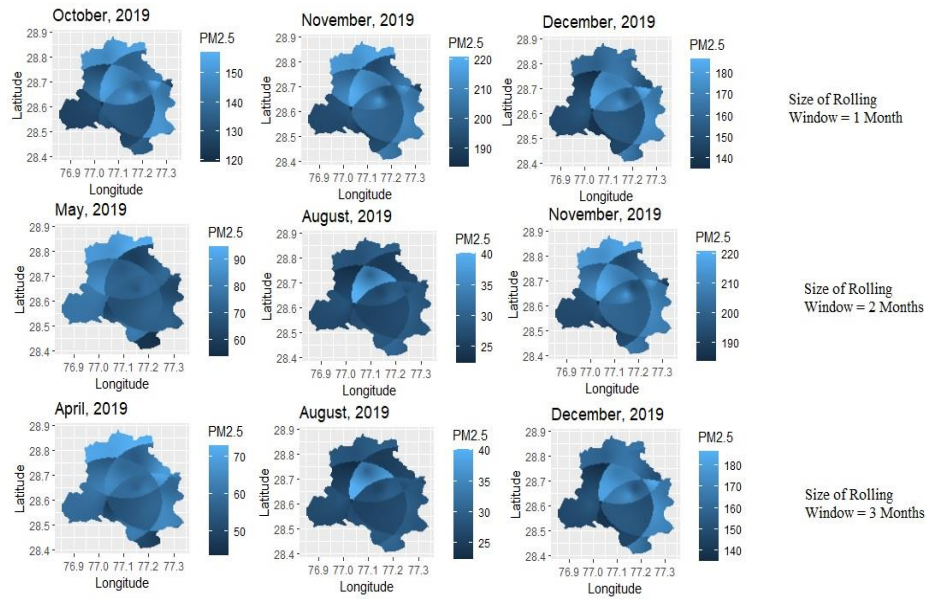

**Fig. 2.** Hybrid Spatio-temporal copula interpolation for different rolling window size.

Though, this algorithm is advantageous compared to others having some serious limitations. In this research, the assumption of isotropic and intrinsic stationarity of spatial dependence is hardly achieved by ignoring the effect of spatial boundary points. Therefore, there are some scopes to improve this research.

## 5   Conclusions

The proposed algorithm performs with significant accuracy. This research work distinguishes spatial trends and temporal dependency as well. Instead of handling nonstationary random fields, we emphasize local stationarity ignoring Gaussian assumption. The possible advancement is tackling Spatio-Temporal dependence via copula and employment of transfer learning in BLSTM architecture.